\documentstyle[12pt,epsfig,amstex]{article}
\begin{document}
%
%

\def\etal{{\em et al.}}

\newcommand{\mv}{\,\rm{mV}}
\newcommand{\V}{\,\rm{V}}
\newcommand{\mG}{\,\rm{mG}}
\newcommand{\nm}{\,\rm{nm}}
\newcommand{\mm}{\,\rm{mm}}
\newcommand{\cm}{\,\rm{cm}}
\newcommand{\m}{\,\rm{m}}
\newcommand{\km}{\,\rm{km}}
\newcommand{\mpc}{\,\rm{Mpc}}
\newcommand{\kel}{\,^0K}
\newcommand{\sr}{\,\rm{sr}}
\newcommand{\pmt}{\it PMT}
\newcommand{\phe}{\it phe}
\newcommand{\y}{\,\rm{y}}
\newcommand{\days}{\,\rm{d}}
\newcommand{\hou}{\,\rm{h}}
\newcommand{\minu}{\,\rm{m}}
\newcommand{\s}{\,\rm{s}}
\newcommand{\usec}{\,\mu{\rm s}}
\newcommand{\msec}{\,\rm{ms}}
\newcommand{\nsec}{\,\rm{ns}}
\newcommand{\psec}{\,\rm{ps}}
\newcommand{\hz}{\,\rm{Hz}}
\newcommand{\khz}{\,\rm{kHz}}
\newcommand{\Mhz}{\,\rm{MHz}}
\newcommand{\mg}{\,\rm{mg}}
\newcommand{\lit}{\,\rm{l}}
\newcommand{\gra}{\,\rm{g}}
\newcommand{\ev}{\,\rm{eV}}
\newcommand{\kev}{\,\rm{KeV}}
\newcommand{\mev}{\,\rm{MeV}}
\newcommand{\gev}{\,\rm{GeV}}
\newcommand{\mw}{\,\rm{MW}}
\newcommand{\gw}{\,\rm{GW}}
\newcommand{\celsi}{\,^o\rm{C}}
\newcommand{\ele}{\rm e^-}
\newcommand{\pos}{\rm e^+}
\newcommand{\pro}{\rm p}
\newcommand{\neu}{\rm n}
\newcommand{\deu}{\rm d}
\newcommand{\nue}{\nu_e}
\newcommand{\anue}{\bar{\nu}_e}
\newcommand{\numu}{\nu_{\mu}}
\newcommand{\anumu}{\bar{\nu}_{\mu}}
\newcommand{\nutau}{\nu_{\tau}}
\newcommand{\anutau}{\bar{\nu}_{\tau}}
\newcommand{\nux}{\nu_x}
\newcommand{\anux}{\bar{\nu}_x}
\newcommand{\anu}{\bar{\nu}}
\newcommand{\vect}[1]{\overrightarrow{\sf #1}}
\newcommand{\system}[1]{\left\{\matrix{#1}\right.}
\newcommand{\displayfrac}[2]{\frac{\displaystyle #1}{\displaystyle #2}}
\newcommand{\nucl}[2]{{}^{#1}\mbox{#2}}
\newcommand{\diff}{{\rm\,d}}
\newcommand{\lsim}{\,\lower .5ex\hbox{$\buildrel < \over {\sim}$}\,}
\newcommand{\gsim}{\,\lower .5ex\hbox{$\buildrel > \over {\sim}$}\,}
\newcommand{\apm}[2]{\,\lower .5ex \buildrel #1 \over #2 \,}
\newcommand{\fact}[1]{#1{\mbox !}}
\newcommand{\varr}[2]{\lb\begin{array}{c} #1 \\ #2 \end{array}\rb}
\newcommand{\bd}[1]{\mbox{\boldmath{${#1}$}}}
\newcommand{\alf}[1]{\mbox{#1}}
\newcommand{\ra}{\rangle}
\newcommand{\la}{\langle}
\newcommand{\lra}{\longrightarrow}
\newcommand{\rb}{\right)}
\newcommand{\lb}{\left(}
\newcommand{\be}{\begin{equation}}
\newcommand{\ee}{\end{equation}}
\newcommand{\bn}{\bd{\nabla}}
\newcommand{\ii}{\mbox{i}}
\newcommand{\betm}{\beta^-}
\newcommand{\betp}{\beta^+}

\def\dmsq{\delta m^{2}}
\def\sinsq{{\rm sin}^2 2\theta}
\def\Reines{$\overline{\nu}_e\,+\,p\,\rightarrow\,e^+\,+\,n$}

\def\isotope#1{\mbox{${}^{#1}$}}                  
\def\ra{\rightarrow}
\def\lra{\leftrightarrow}
\def\units#1{\hbox{$\,{\rm #1}$}}                
%
\begin{center}
{\bf Limits on Neutrino Oscillations from the CHOOZ  Experiment}
\end{center}
\begin{center}
M.~Apollonio$^c$,
A.~Baldini$^b$,
C.~Bemporad$^b$,
E.~Caffau$^c$,
F.~Cei$^b$,
Y.~D\'eclais$^{e,1}$,
H.~de~Kerret$^f$,
B.~Dieterle$^h$,
A.~Etenko$^d$,
J.~George$^h$,
G.~Giannini$^c$,
M.~Grassi$^b$,
Y.~Kozlov$^d$,
W.~Kropp$^g$,
D.~Kryn$^f$,
M.~Laiman$^e$,
C.E.~Lane$^a$,
B.~Lefi\`evre$^f$,
I.~Machulin$^d$,
A.~Martemyanov$^d$,
V.~Martemyanov$^d$,
L.~Mikaelyan$^d$,
D.~Nicol\`o$^b$,
M.~Obolensky$^f$,
R.~Pazzi$^b$,
G.~Pieri$^b$,
L.~Price$^g$,
S.~Riley$^g$,
R.~Reeder$^h$,
A.~Sabelnikov$^d$,
G.~Santin$^c$,
M.~Skorokhvatov$^d$,
H.~Sobel$^g$,
J.~Steele$^a$,
R.~Steinberg$^a$,
S.~Sukhotin$^d$,
S.~Tomshaw$^a$,
D.~Veron$^f$,
and V.~Vyrodov$^f$
\end{center}
\begin{center}
$^a${\em   Drexel University                    }  \\
$^b${\em   INFN and University of Pisa          }  \\
$^c${\em   INFN and University of Trieste       }  \\
$^d${\em   Kurchatov Institute                  }  \\
$^e${\em   LAPP-IN2P3-CNRS Annecy               }  \\
$^f${\em   PCC-IN2P3-CNRS Coll\`ege de France   }  \\
$^g${\em   University of California, Irvine     }  \\
$^h${\em   University of New Mexico, Albuquerque}  \\
$^1${\em   Present address: IPNL-IN2P3-CNRS Lyon}  \\
\end{center}
%
%
\begin{abstract}
\noindent
We present new results based on the entire CHOOZ\footnote{%
The CHOOZ experiment is named after the new nuclear power station
operated by \'Electricit\'e de France (EdF) near the village of Chooz in
the Ardennes region of France.}
data sample. 
We find (at 90\,\% confidence level) no evidence for neutrino oscillations 
in the $\anue$ disappearance mode, for the parameter region given 
by approximately  $ \dmsq > 7 \cdot 10^{-4}\units{eV^2} $ for maximum mixing, 
and $\sinsq = 0.10$ for large $\dmsq$. Lower sensitivity results, 
 based only on the comparison of the positron spectra from the two
different-distance nuclear reactors, are also presented; these are 
independent of the absolute normalization of the $\anue$ flux, the cross 
section, the number of target protons and the detector efficiencies.
\end{abstract}


\section{Introduction}
Preliminary results of the CHOOZ experiment have already been 
 published\cite{Apo98}. We  present here the  new  results based on the entire 
data sample; they include a large increase in statistics 
and  a better understanding of systematic effects. 

The reader is refered to the previous article for an introduction to the problem of 
neutrino oscillations, for a general description of the experiment and 
for a discussion of its data analysis.

As the experiment progressed, calibration methods and stability checks were 
considerably refined, and knowledge of the apparatus' 
behaviour and   simulation by the Montecarlo method were improved
As a consequence, systematic errors were considerably reduced.

Three results are given. The main one is based on all the available 
information: the measured number of positron events as a function of 
 energy, separately obtained 
from each reactor. 
It uses the two spectral shapes, as well as the absolute 
normalizations. The second result is based only on the comparison of the 
positron spectra from the two, different-distance  reactors.
This analysis is largely unaffected by the absolute value of the $\anue$ 
flux, the cross section, the number of target protons and the detector 
efficiencies, and is therefore dominated by statistical errors. 
The sensitivity in this case is limited to 
$\delta m^2 \gsim 2 \cdot 10^{-3} \units{eV^2}$ 
due to the small distance,  $\Delta L = 116.7$ m, between the reactors.
The explored $(\delta m^2,\sin^2 2 \theta)$ parameter space still matches 
well the region of the atmospheric neutrino anomaly.
The third analysis is similar to the first, but does not include the 
absolute normializations.
All results were derived following the suggestions by 
Feldman \& Cousins~\cite{Fel98} \footnote{The previous results
\cite{Apo98}, were published before the unified statistical approach was 
proposed~\cite{Fel98}; they excluded therefore a slightly larger 
parameter region}.   

\section{Experimental data}
\label{sec:data_taken}
The Chooz power station has two pressurized-water reactors with a total
thermal power of $8.5 \,{\gw}_{th}$. The first reactor reached full
power in May $1997$, the second in August $1997$.
A summary of our data taking (from April 7, 1997 to July
20, 1997) is presented in Table~\ref{dataacq}.

\begin{table}[htbp]
\caption{\small Summary of the Chooz data acquisition cycle
         from April $1997$ to July $1998$.}
\label{dataacq}
\begin{center}
\begin{tabular}{lcc}
\hline
                    & Time (h)  & $\int W \diff t$ (\gw h)\\
\hline
Run                 & 8761.7    & \\
Live time           & 8209.3    & \\
Dead time           & 552.4     & \\
Reactor 1 only ON   & 2058.0    & 8295 \\
Reactor 2 only ON   & 1187.8    & 4136 \\
Reactors 1 \& 2 ON  & 1543.1    & 8841 \\
Reactors 1 \& 2 OFF & 3420.4    & \\
\hline
\end{tabular}
\end{center}
\end{table}
Note  that the schedule was quite convenient for 
separating the individual reactor contributions and for determining the 
reactor-OFF background.

In this exepiment the $\anue$'s are detected via 
the inverse $\beta$-decay reaction
%
$$ \anue + \pro \rightarrow \pos + \neu $$
\noindent
The $\anue$ reaction signature is a delayed coincidence between the prompt
$\pos$ signal (boosted by the two 511-\units{keV} annihilation $\gamma$ rays),
later called ``primary signal'', and the signal due to the neutron capture 
in the Gd-loaded scintillator ($\gamma$-ray energy release $\sim 8 
\units{MeV}$), later called ``secondary signal''. During the experiment 
$1.2 \times 10^7$ events were recorded on disk; weak selection criteria, 
based on the total charge measured by the PMT's for the secondary signal, 
reduced this number to $7 \times 10^5$ events, which were fully reconstructed 
in energy and in space. After applying the criteria for selecting 
$\nu$-interactions, we were left with $2991$ {\it bona-fide} candidates, 
including  $287$ events from reactor-OFF periods.

\subsection{Detector stability}
\begin{figure}[htb]
\begin{center}
\mbox{\epsfig{file=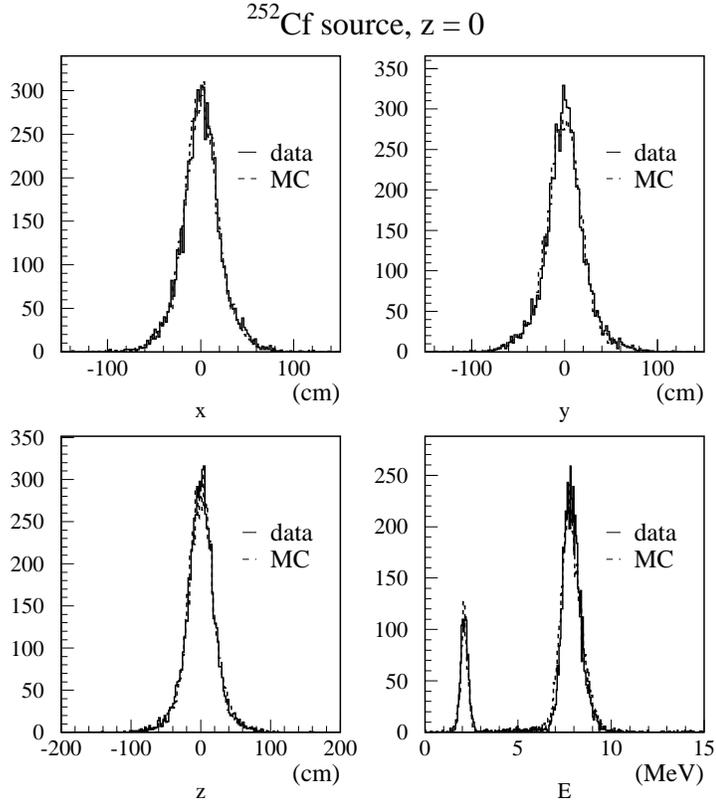,width=0.75\textwidth}}
\caption{\small Visible energy and position distributions of 
$^{252}$Cf source calibration data (at the detector center): 
comparison between data and Monte Carlo simulation.}
\label{calibr}
\end{center}
\end{figure}

During our approximately one year of data taking, the detector slowly 
varied its response due to the decrease of the optical clarity of the 
Gd-loaded scintillator (the scintillator emission was stable, but the 
light at the PMT's exponentially decreased, with a time constant   
$\tau \sim 750 \units{d}$). This produced small effects on the trigger 
threshold and rate, the event reconstruction, the signal/background 
separation and the background level. While hardware thresholds were 
readjusted every few months, the detector response was  checked daily 
by $\nucl{60}{Co}$, $\nucl{252}{Cf}$ and Am/Be sources, which 
provide $\gamma$-signals, neutron signals, and time correlated 
$\gamma - \neu$ signals. The reconstruction of these event samples, the 
study of their time evolution, and the comparison with Montecarlo method 
predictions, permitted a thorough understanding of the detector behaviour  
and a precise evaluation of the small efficiency variations on 
neutrino-induced and background events.  
Figure \ref{calibr} shows calibration data using the $\nucl{252}{Cf}$ 
source at the detector center; the neutron capture lines ($2.2 \units{MeV}$ 
on hydrogen and $8 \units{MeV}$ on gadolinium) are compared with Montecarlo 
 predictions. Position and energy resolutions are 
$\sigma_x = 17.5 \units{cm}$ and $\sigma_E/E = 5.6 \,\%$ for n-captures 
releasing $8 \units{MeV}$.  Calibrations at other locations always
 produced detector 
response  in good agreement with the  Montecarlo  predictions. 

Interestingly, we are sensitive to the two-line structure of the gadolinium capture 
at $8 \units{MeV}$. A fit to the data gives  line  energies and 
intensities of $7.77 \units{MeV} \quad (77 \,\%)$ for $\nucl{157}{Gd}$ and 
$8.31 \units{MeV} \quad (23 \,\%)$ for $\nucl{155}{Gd}$. The quality of 
the fit is good ($\chi^2 = 67.6$ with  55 dof); a single-Gaussian fit 
gives a much poorer result ($\chi^2 = 875$ with 58 dof). 

\begin{figure}[htb]
\begin{center}
\mbox{\epsfig{file=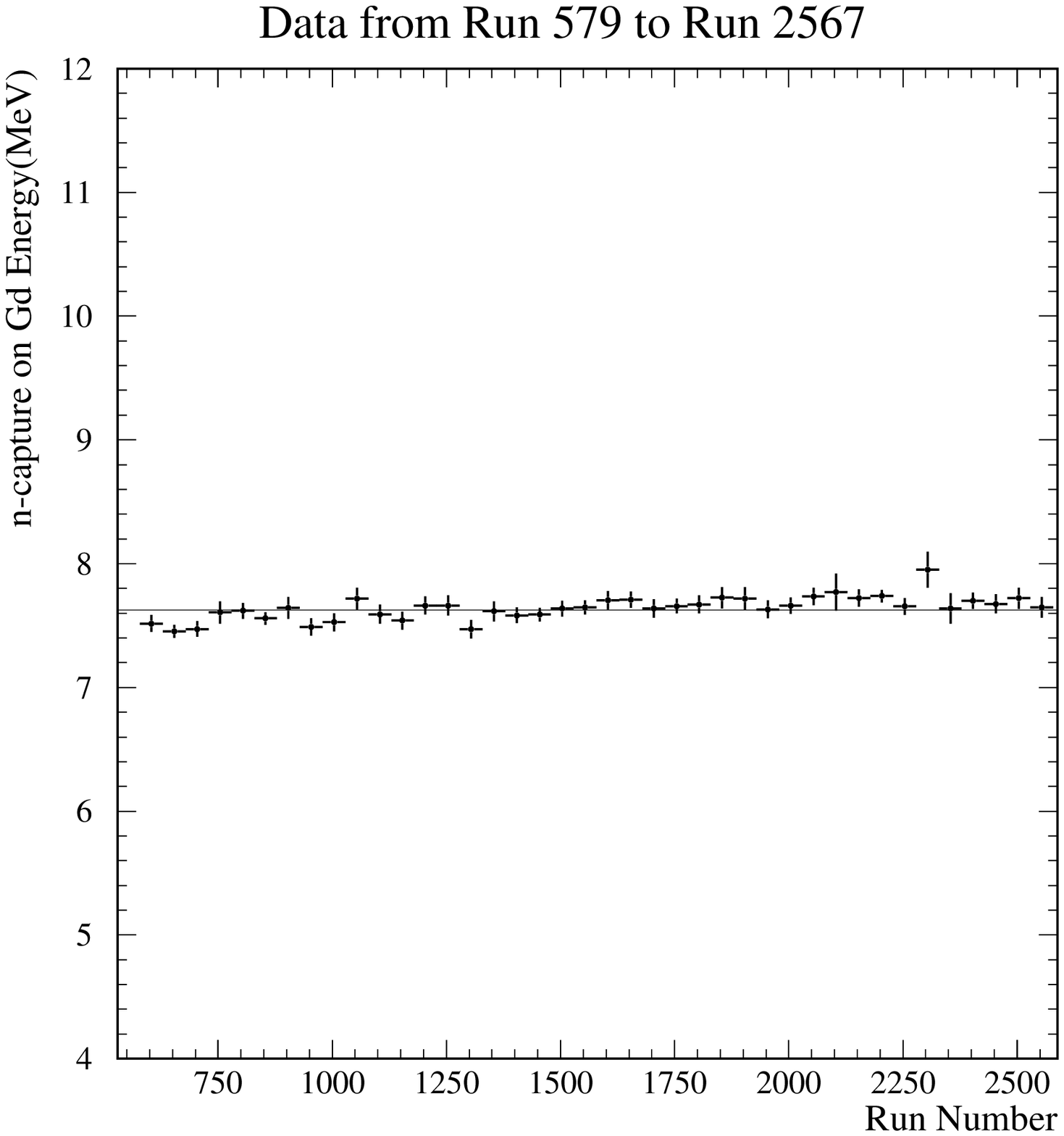,width=0.45\textwidth}
      \epsfig{file=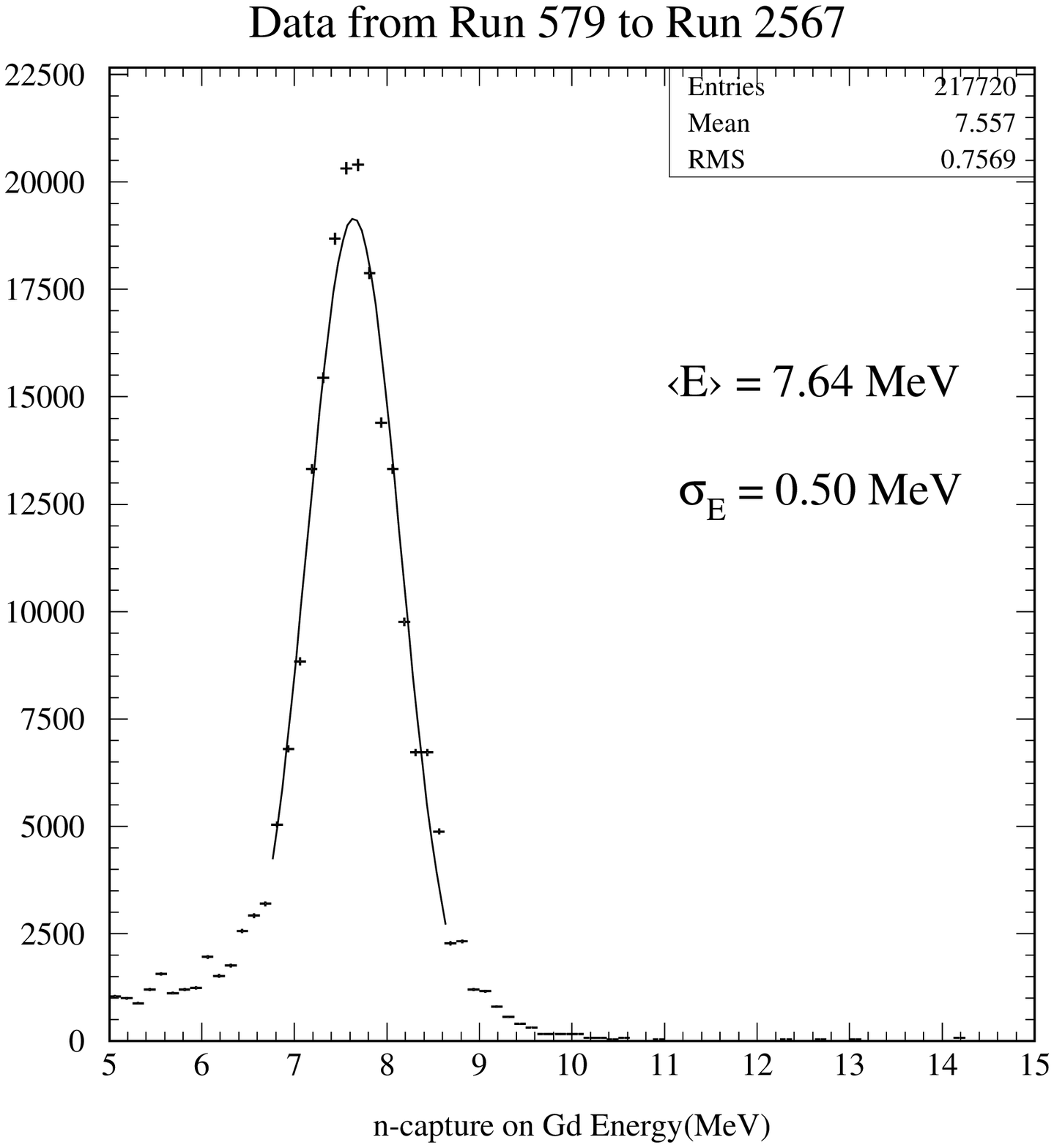,width=0.45\textwidth}
      }
\caption{\small Position of the peak (left) and lineshape (right) 
of the $8 \units{MeV}$ $\gamma$-line.}
\label{stab}
\end{center}
\end{figure}

As a demonstration of the excellent 
stability of the detector response, Fig. 
\ref{stab} shows  the time  evolution of the measured energy 
correponding to the $8 \units{MeV}$ capture line and the shape and width 
of this line, for spallation neutrons generated by cosmic ray muons 
during the entire duration of the experiment. Since these events occured 
everywhere in the detector, the data in  Fig. \ref{stab} depends on daily 
calibrations, on the determination of all PMT and electronic channel 
amplification constants, on the knowledge of the scintillator attenuation 
length and its time evolution, and on the event reconstruction algorithms. 
The measured energy is somewhat lower than $8 \units{MeV}$, due to  
scintillator saturation effects and 
neutron-capture $\gamma$-ray leakage. 
%
\subsection{Event Typology}
\begin{figure}[htbp]
\begin{center}
\mbox{\epsfig{file=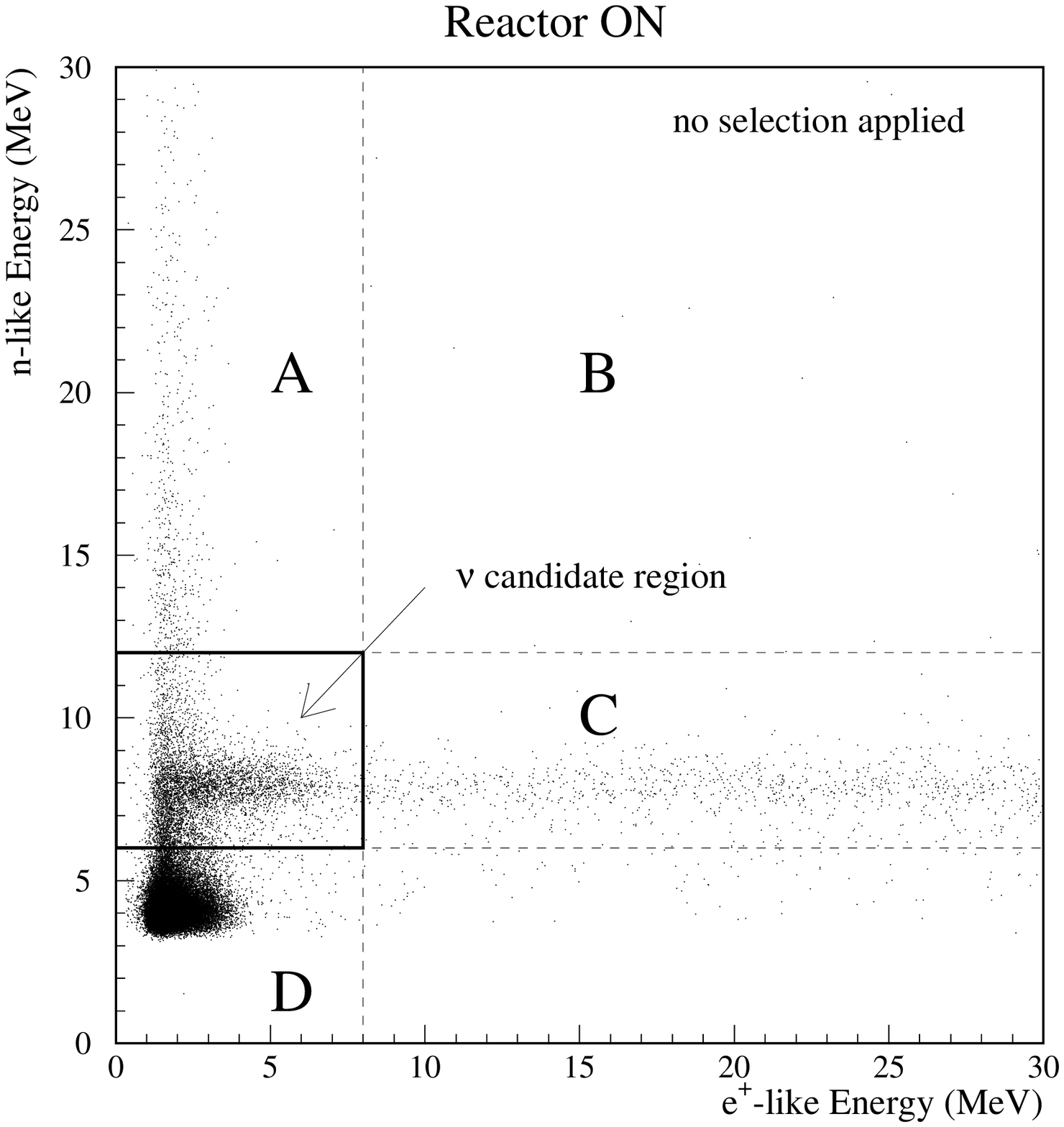,width=0.5\textwidth}
      \epsfig{file=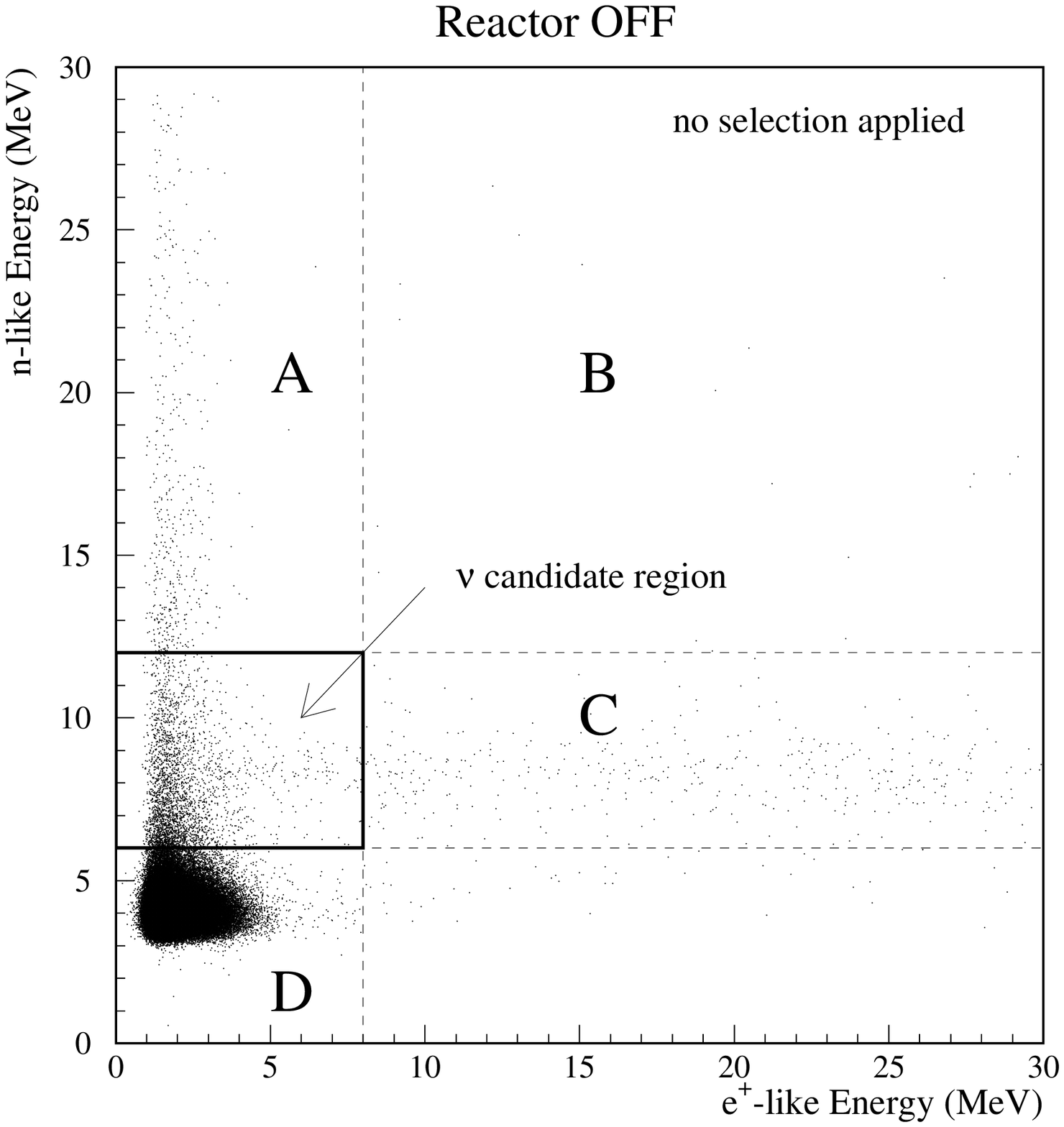,width=0.5\textwidth}}
\caption{\small ``n-signal energy'' vs. ``$\pos$-signal energy'' during the 
reactor-ON (left) and reactor-OFF (right) period; no selection applied.}
\label{reacon}
\end{center}
\end{figure}

\begin{figure}[htbp]
\begin{center}
\mbox{\epsfig{file=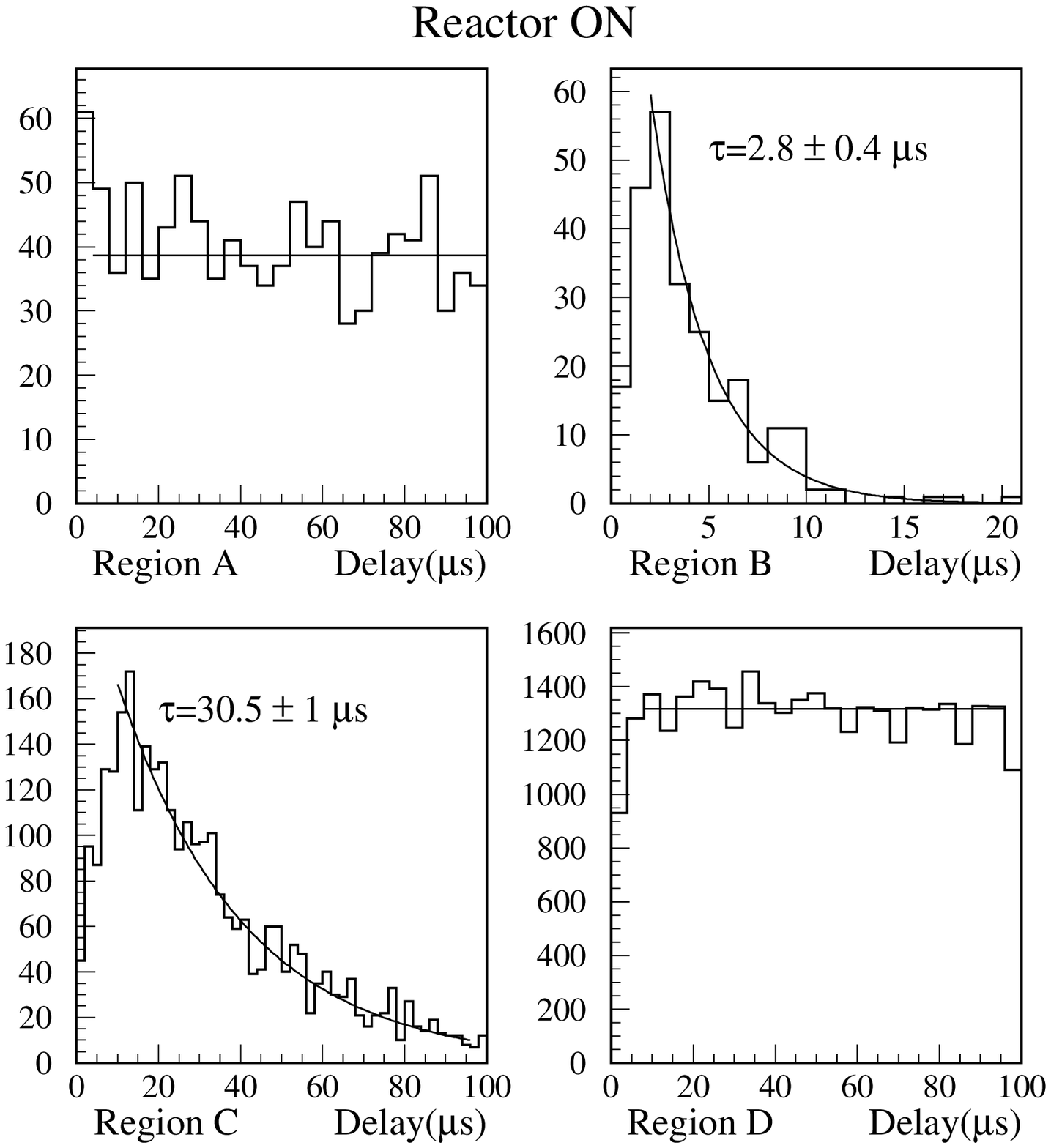,width=0.75\textwidth}}
\vspace{-2.2cm}
\caption{\small Delay distribution of secondary events 
in the various regions.}
\label{delay}
\end{center}
\end{figure}

A good understanding of the  nature of the  neutrino candidates can be 
obtained by viewing the events on a two-dimensional plot of 
``n-signal energy'' vs. ``$\pos$-signal energy'', for reactor-ON (Fig. 
\ref{reacon} (left)) and reactor-OFF (Fig. \ref{reacon} (right)) data; 
no signal selection has yet  been applied.
%
%
One can observe four regions: A,B,C,D and the neutrino event window at 
the crossing of regions A and C. Regions B and C are filled by 
primary-secondary correlated signals. Region B contains stopping 
muons, i.e.: cosmic $\mu$'s which entered the detector through 
the small dead space (detector filling pipes, support flanges, etc.) 
missing the anticoincidence shield. These events have large primary 
energy and large secondary energies associated with the $\mu$-decay 
electrons. Events in region B have a secondary delay distribution 
in agreement with the $\mu$-lifetime at rest (see Fig. \ref{delay}B).

Region C events are due to fast neutrons from nuclear spallations by 
cosmic rays in the rock and concrete surrounding the detector; these 
neutrons scatter and the recoil proton is detected as ``primary'', 
while the neutron is thermalized and later captured as the 
``secondary'' giving the characteristic $8 \units{MeV}$ capture energy; 
this fast neutron region overlaps the neutrino candidate region and 
is the main background source for the experiment. The secondary delay 
distribution is as expected for thermal neutron capture in the 
Gd-doped scintillator (the best-fit lifetime is $\tau=30.5\, \pm \, 
1 \units{\mu s}$) (see Fig. \ref{delay}C). Regions A and D are filled by  
accidental events; region D events are due to the accidental coincidence 
(within $100 \units{\mu s}$) of two low energy natural radioactivity 
signals; region A events are due to an accidental coincidence of a low 
energy natural radioactivity signal and a high energy recoil proton from 
a fast neutron scattering. Both A and D delay distributions are flat, 
as expected (see Fig.~\ref{delay}A and D).

\begin{figure}[htbp]
\begin{center}
\mbox{\epsfig{file=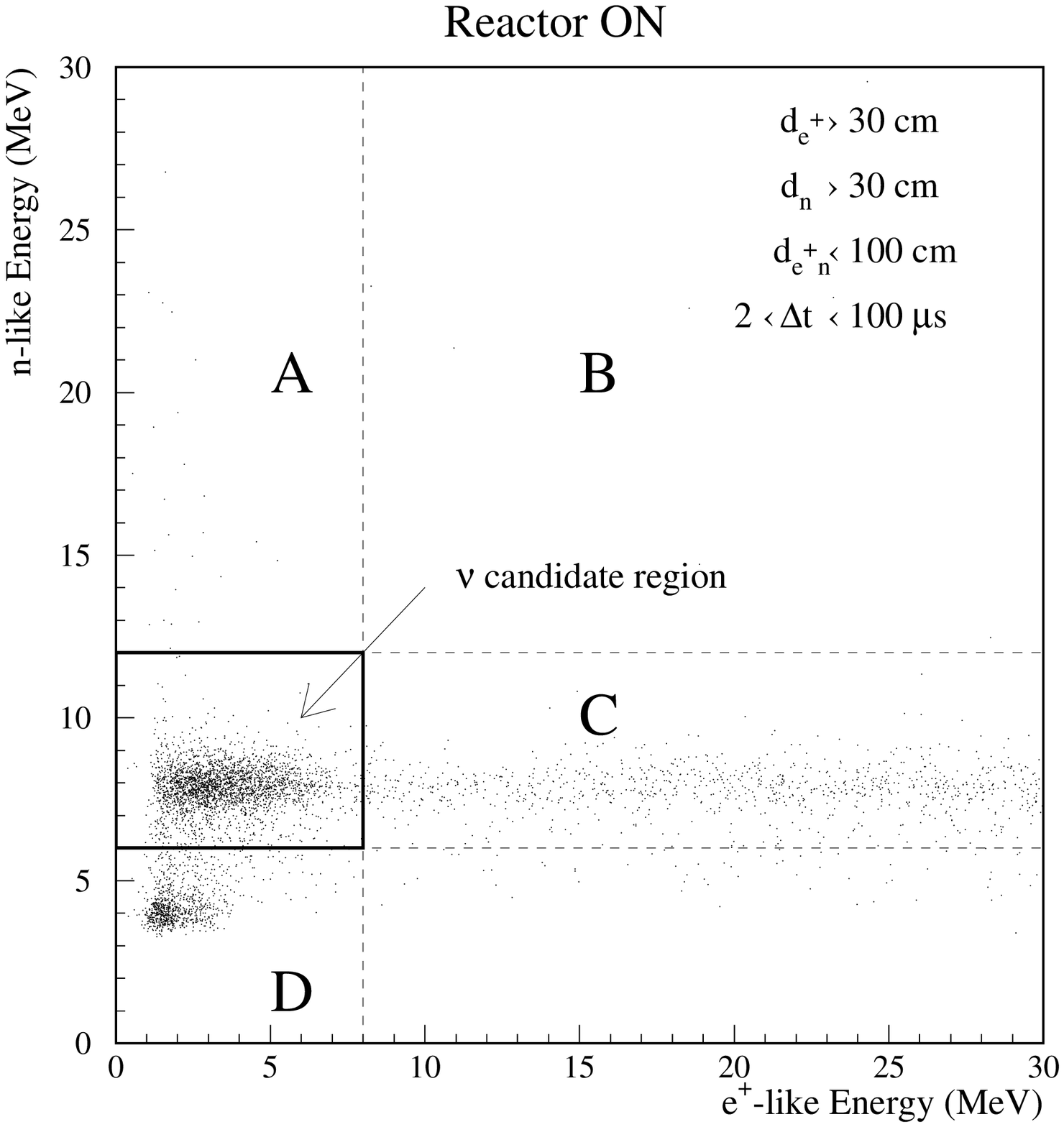,width=0.5\textwidth}
     \epsfig{file=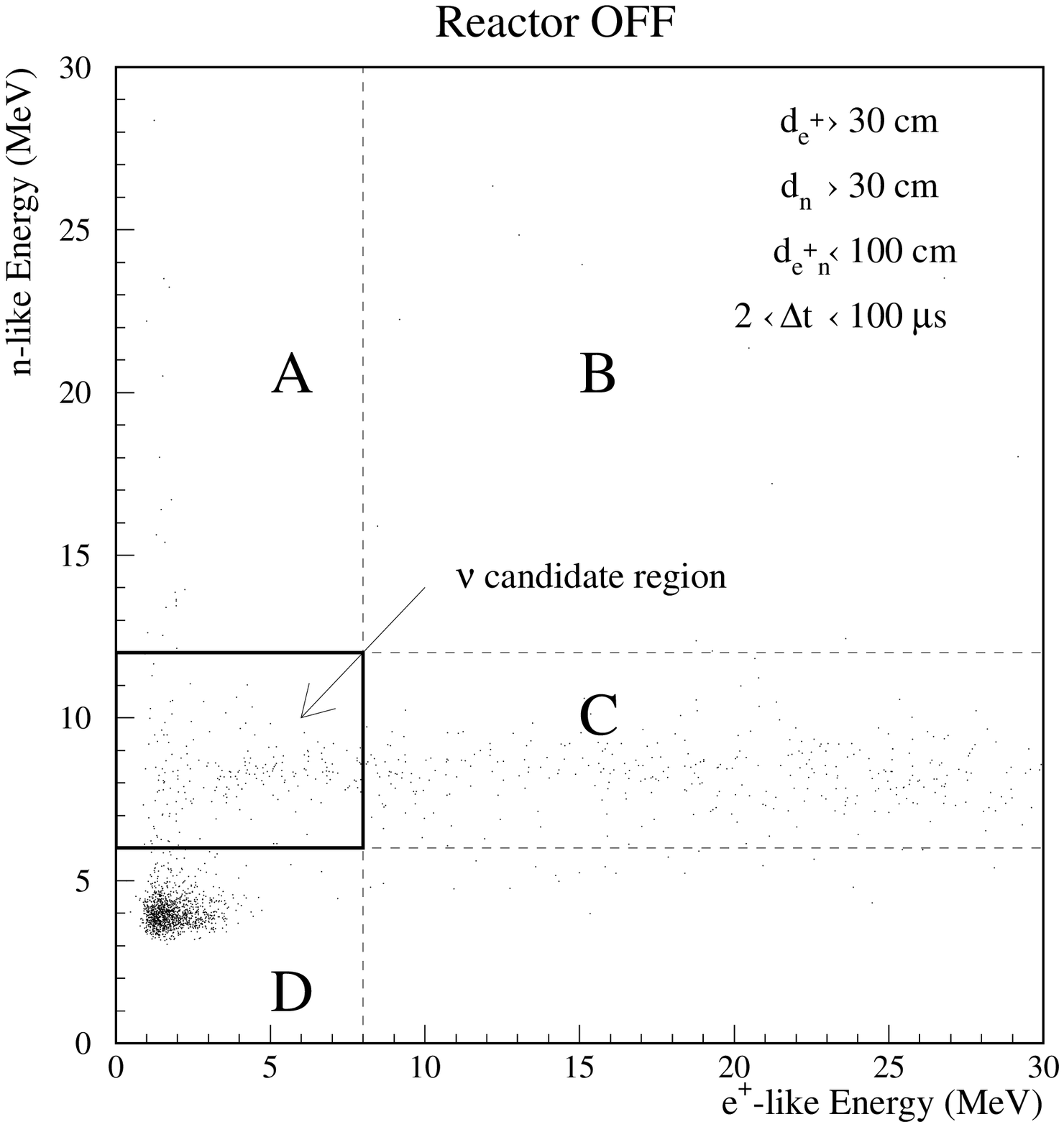,width=0.5\textwidth}
      }
\caption{\small ``n-signal energy'' vs. ``$\pos$-signal energy'' for 
reactor-ON (left) and reactor-OFF (right) data; selections other than 
energy applied.}
\label{reaconcut}
\end{center}
\end{figure}

 The definition of a neutrino 
event is based on the following requirements:
\begin{itemize}
\vspace{-1mm}
\item energy cuts on the neutron candidate ($6 - 12 \units{MeV}$) and 
on the $\pos$ candidate (from the threshold energy $E_{thr} \sim 1.3 
\units{MeV}$ to $8 \units{MeV}$),
\vspace{-1mm}
\item a time window on the delay between the $e^+$ and the
neutron ($2 - 100 \units{\mu s}$), 
\vspace{-1mm}
\item spatial selections on the $\pos$ and the neutron positions 
(distance from the PMT wall $> 30 \units{cm}$ and distance 
between $\neu$ and $\pos < 100 \units{cm}$),
\vspace{-1mm}
\item only one pulse satisfying the criteria for a secondary signal
(neutron).
\end{itemize}
The application of these selection criteria (apart from energy selections) 
produces the two-dimensional plots of Fig. \ref{reaconcut}, for reactor-ON 
(left) and for reactor-OFF (right) data.
%
%
One can clearly see that the events spilling into the neutrino event 
window are mainly those from region C (proton recoils and neutron 
capture from spallation fast neutrons) and, to a lesser extent, those 
from region D (two low energy natural radioactivity signals). 

The effects of the selection criteria used to define the neutrino 
interactions were extensively studied by the Montecarlo method and 
by {\it ad-hoc} $\gamma$ and $n$-source calibrations. Similarly, we 
investigated the small edge effects associated with the acrylic vessel 
containing the Gd-loaded scintillator target. In Table \ref{effi} 
we present the efficiencies associated with the selection criteria 
and their errors. 
\begin{table}[htbp]
  \caption{\small Summary of the neutrino detection efficiencies.}
  \label{effi}
  \begin{center}
    \begin{tabular}{lcc}
      \hline
      selection                   & efficiency $(\%)$ & error $(\%)$ \\
      \hline
      positron energy             & $97.8$            & $0.8$ \\
      positron-geode distance     & $99.85$           & $0.1$ \\
      neutron capture             & $84.6$            & $0.85$ \\
      capture energy containment  & $94.6$            & $0.4$ \\
      neutron-geode distance      & $99.5$            & $0.1$ \\
      neutron delay               & $93.7$            & $0.4$ \\
      positron-neutron distance   & $98.4$            & $0.3$ \\
      secondary multiplicity      & $97.4$            & $0.5$ \\
      combined                    & $69.8$            & $1.1$ \\
      \hline
    \end{tabular}
  \end{center}
\end{table}
%

\subsection{Positron spectrum}
\begin{figure}[htb]
\begin{center}
\mbox{\epsfig{file=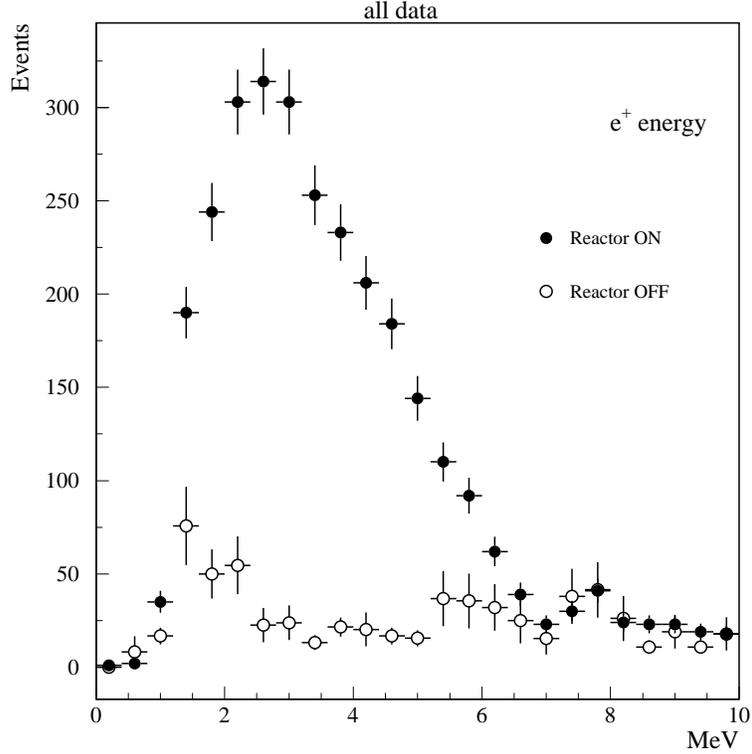,width=0.75\textwidth}}
\caption{\small Positron energy spectra in reactor-ON and OFF periods.}
\label{posspe}
\end{center}
\end{figure}

\begin{figure}[hp]
  \begin{center}
    \mbox{\epsfig{file=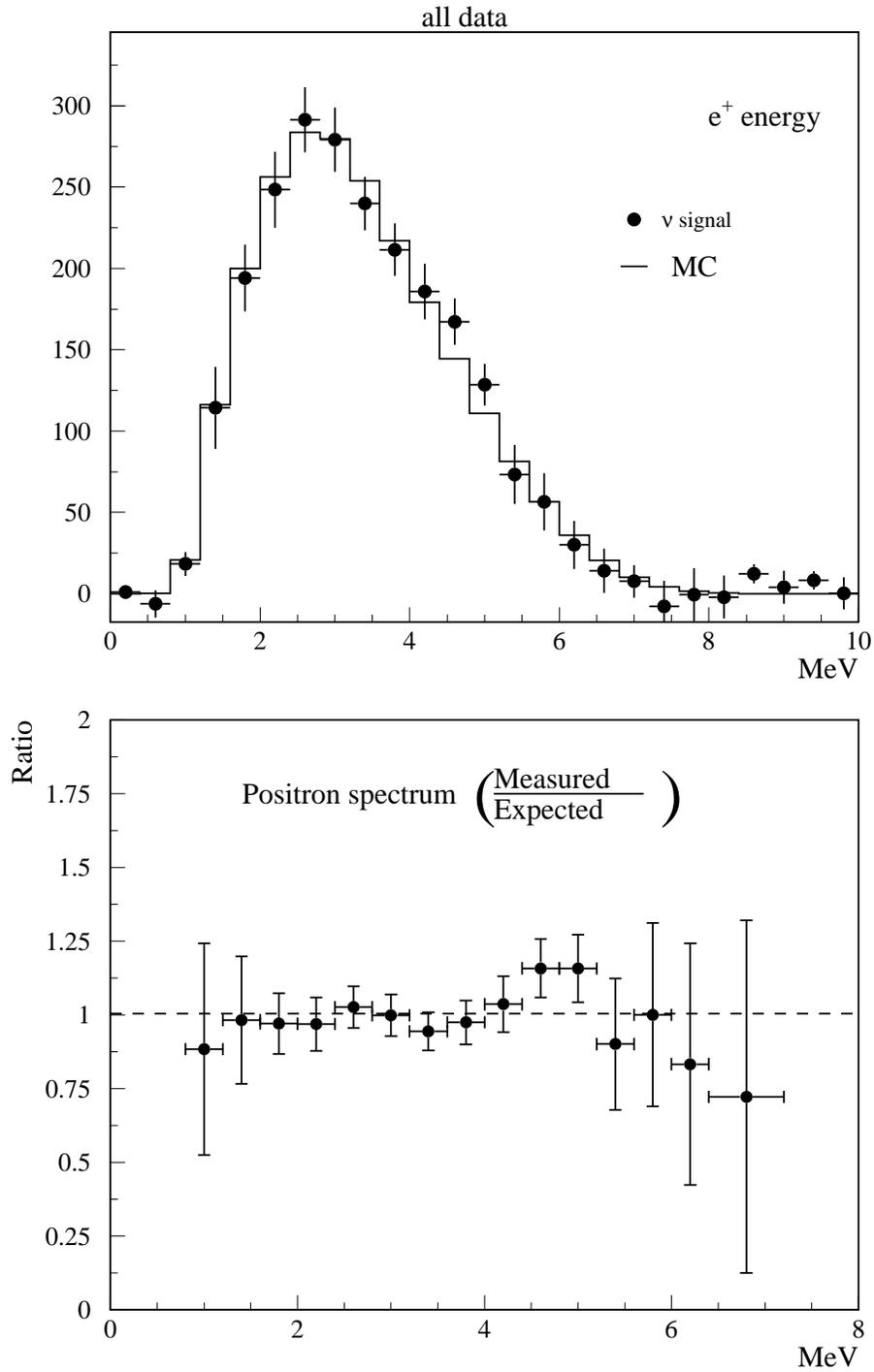,width=0.9\linewidth}}
    \caption{\small (above) Expected positron spectrum for the case of no 
                    oscillations, superimposed on the measured positron 
                    spectrum obtained from the subtraction of reactor-ON 
                    and reactor-OFF spectra; (below) measured vs. expected 
                    ratio. The errors shown are statistical.}
    \label{posonoff}
  \end{center}
\end{figure}

The measured positron spectrum for all reactor-ON data, 
and the corresponding reactor-OFF spectrum, are shown in Fig. 
\ref{posspe}.
After background subtraction, the measured positron spectrum can be 
compared with the expected neutrino-oscillated  
positron spectrum at the detector position. For a mean reactor-detector 
distance $L_k$, this is given by:
\begin{equation}
  \begin{split}
    S_k(E,L_k,\theta,\delta m^2) &= \frac{1}{4\pi L_k^2} n_p
    \int h(L,L_k)  \int \sigma(E_\nu) S(E_\nu)\times \\ 
    &\quad P(E_\nu,L,\theta,\delta m^2) r(E_{e^+},E) \varepsilon(E_{e^+})
    \diff E_{e^+} \diff L,
  \end{split}
  \label{posexp}
\end{equation}
where 
\begin{center}
\begin{tabular}{ll}
$E_\nu,E_{e^+}$                & are related by $E_{\nu} = E_{e^+} 
                                 + (M_n - M_p) + O(E_\nu/M_n)$, \\
$n_p$                          & is the total number of target protons, \\
$\sigma(E_\nu)$                & is the neutrino cross section, \\
$S(E_\nu)$                     & is the antineutrino spectrum, \\
$h(L,L_k)$                     & is the spatial distribution 
                                 function for the \\
                               & finite core and detector sizes, \\ 
$r(E_{e^+},E)$                 & is the detector response function linking \\
                               & the visible energy $E$ and the real 
                                 positron energy $E_{e^+}$, \\
$\varepsilon(E_{e^+})$          & is the neutrino detection efficiency, \\
$P(E_\nu,L,\theta,\delta m^2)$ & is the two-flavour survival probability.
\end{tabular}
\end{center}
%
%
The $\anue$ spectrum was determined, for each fissile isotope, by using 
the $\anue$ yields obtained by conversion of the $\beta^{-}$-spectra 
measured at ILL~\cite{Sch85}; these spectra were then renormalized 
according to the measurement of the integral $\anue$ flux performed at 
Bugey\cite{Dec94}.  
The expected, non-oscillated positron spectrum was computed using the Monte Carlo codes to simulate both reactors and
the detector. The resulting spectrum, summed over the two reactors, is 
superimposed on the measured one in Fig.~\ref{posonoff} to emphasize 
the agreement of the data with the no-oscillation hypothesis. The 
Kolmogorov-Smirnov test for the compatibility of the two distributions 
gives an $82\,\%$ probability. 
The measured vs. expected ratio, averaged over the energy spectrum 
(also presented in Fig.~\ref{posonoff}) is
\begin{equation}
R = 1.01 \pm 2.8 \,\% ({\rm stat}) \pm 2.7 \,\% ({\rm syst})
\end{equation}
\subsection{Neutrino interaction yield}
As shown in Table \ref{dataacq}, we collected data during 
reactor-OFF periods and periods of power rise for each reactor. 
This had two beneficial consequences: first, the collection of enough 
reactor-OFF data to precisely determine the amount of background; 
second, the measurement of the neutrino interaction yield as a 
function of the reactor power. By fitting the slope of the measured 
yield versus reactor power, one can obtain an estimate of the neutrino 
interaction yield at full power, which can then be compared with 
expectations and with the oscillation hypothesis.

 The fitting procedure 
is carried out as follows. For each run the expected number of neutrino 
candidates results from the sum of a signal term, linearly dependent 
on the reactor power, and a background term, assumed to be constant 
and independent of power. Thus
\begin{equation}
  \overline{N}_i = (B + W_{1i}Y_{1i} + W_{2i}Y_{2i}) \Delta t_i,
  \label{nrun}
\end{equation}
where the index $i$ labels the run number, $\Delta t_i$ is the corresponding
live time, $B$ is the background rate, ($W_{1i},W_{2i})$ are the thermal 
powers of the two reactors in GW and $(Y_{1i},Y_{2i})$ the positron yields 
per GW induced 
by each reactor. These yields still depend on the reactor index (even in the 
absence of neutrino oscillations), because of the different distances, and 
on run number, as a consequence of their different and varying fissile 
isotope compositions. It is thus convenient to factorize $Y_{ki}$ into a 
function $X_k$ (common to both reactors in the no-oscillations case) and 
distance dependent terms, as follows:
\begin{equation}
  Y_{ki} = (1 + \eta_{ki})\frac{L_1^2}{L_k^2} X_k,
  \label{nfact}
\end{equation}
where $k = 1,2$ labels the reactors and the $\eta_{ki}$ corrections contain 
the dependence of the neutrino interaction yield on the fissile isotope 
composition of the reactor core and the positron efficiency corrections.
We are thus led to define a cumulative ``effective'' power according to 
the expression%
\footnote{The ``effective'' power may be conceived as the thermal power 
released by a one-reactor station located at the reactor 1 site, 
providing $9.55 \gw$ at full operating conditions and at starting of reactor 
operation (no burn-up).}
\begin{equation}
  W_i^\ast \equiv \sum_{k = 1}^2 W_{ki} (1 + \eta_{ki})\frac{L_1^2}{L_k^2} \ . 
  \label{weff}
\end{equation}
Eqn.(\ref{nrun}) can then be written as 
\begin{equation}
  \overline{N}_i = (B + W_i^\ast\overline{X}) \Delta t_i, 
  \label{nrun2}
\end{equation}
where $\overline{X}$ is the  positron yield per unit power averaged over the two reactors. 
We built the likelihood function ${\cal L}$ by the joint Poissonian 
probability of detecting $N_i$ neutrino candidates when $\overline{N}_i$ 
are expected, and defined
\begin{equation}
  F \equiv -\ln {\cal L} = - \sum_{i=1}^{n} \ln P(N_i;\overline{N}_i) 
  \label{wlike}
\end{equation}
Searching for the maximum likelihood to determine the parameters 
$\overline{X}$ and $B$ is then equivalent to minimizing Eqn. \ref{wlike}.
Both the average positron yield, $\overline{X}$, and the background rate,
 $B$, are assumed to 
be time independent. 

We divided the complete run sample into three periods, according to the 
dates of the threshold resetting (see \S \ref{sec:data_taken}), and 
calculated the fit parameters for each period separately. The results are 
listed in Table~\ref{poilik}.
\begin{table}[htbp]
  \caption{\small Summary of the likelihood fit parameters for the three data 
    taking periods.}
  \label{poilik}
  \begin{center}
    \begin{tabular}{lccc}
      \hline
      period          & 1                      & 2       & 3 \\
      \hline
      starting date   & 97/4/7                 & 97/7/30 & 98/1/12 \\
      runs            & $579 \rightarrow 1074$ & $1082 \rightarrow 1775$ & 
               $1778 \rightarrow 2567$ \\
      live time (h)             & $1831.3$      & $2938.8$ & $3268.4$ \\
      reactor-OFF time (h)      & $38.9$        & $539.5$  & $2737.2$ \\
      $\int W \diff t$ (GWh)    & $7798$        & $10636$  & $2838$\\
      $B$ (counts $\days^{-1}$) & $1.25\pm 0.6$ & $1.22\pm 0.21$  &
             $2.2\pm 0.14$\\
      $\overline{X}$ (counts $\days^{-1}\gw^{-1}$) & $2.60 \pm 0.17$ & $2.60 \pm 0.09$ &
                                          $2.51 \pm 0.17$ \\
      $\chi^2/dof$ & $136/117$ & $135/154$ & $168/184$ \\
      $N_\nu$ (counts $\days^{-1}$ & $24.8 \pm 1.6$ & $24.8 \pm 0.9$ &
                                      $24.0 \pm 1.6$ \\
      \qquad at full power) & & &\\
      \hline
    \end{tabular}
  \end{center}
\end{table}
%
%
The correlated background, evaluated by extrapolating the rate of high 
energy neutrons followed by a capture into the region defined by the 
event selection criteria, turns out to be $1.0 \pm 0.1$ counts 
$\days^{-1}$ for the three data taking periods. We note therefore that 
only the accidental background increased, as expected, following the 
change of the detector response.   
%
By averaging the signal $\overline{X}$ over the three periods, one can obtain
\begin{equation}
  \langle \overline{X} \rangle = (2.58 \pm 0.07) \,\,\,{\rm counts}\days^{-1}\gw^{-1},
  \label{xave}
\end{equation}
corresponding to $(24.7 \pm 0.7)$ daily neutrino interactions at full 
power; the overall statistical uncertainty is  $2.8 \,\%$.

\subsection{Neutrino interaction yield from each reactor}
A similar fitting procedure can be used to determine the contribution to 
the neutrino interaction yield, from each reactor individually,
 and for each energy 
bin of the positron spectra. The generalized Eqn. (\ref{nrun2}) can be 
rewritten in the form:
\begin{equation}
  \overline{N}_i(E_j) = (B(E_j) + W_{1i}^\ast(E_j) X_1(E_j) + 
                                  W_{2i}^\ast(E_j) X_2(E_j))\Delta t_i 
  \label{nruni}
\end{equation}
The spectrum shape is expected to vary, due to the fuel aging (``burnup''),
 throughout the 
reactor cycle. Burnup correction factors $\eta_{ki}$ then need to be 
calculated for each bin of the positron spectrum. The fitted yields, averaged
over the three periods, are listed in Table~\ref{yieboth} and compared to 
the expected yield in the absence of neutrino oscillations.
The yield parameters $X_1,X_2$ are slightly correlated, as shown in 
Table~\ref{yieboth}; such a correlation (which does not exceed $20\,\%$)
is always negative since, at given candidate and background rates, an
 increase
of reactor $1$ yield corresponds to a decrease of reactor $2$ yield (and 
{\it vice versa}). When building the  $\chi^2$ statistic to 
test the oscillation hypothesis, we  take the covariance matrix 
into account.
 
\begin{table}[htb]
  \caption{\small Experimental positron yields for both reactors ($X_1$ and 
           $X_2$) and expected spectrum ($\tilde{X}$) for no-oscillations. 
           The errors ($68 \,\%$ C.L.) and the covariance matrix off-diagonal 
           elements are also listed.}
  \label{yieboth}
  \newcommand{\Rule}{\rule[-.7ex]{0ex}{2.9ex}}
  \begin{center}
    \begin{tabular}{ccccc}
      \hline
      \Rule
      $E_{e^+}$   & $X_1\pm\sigma_1$ & $X_2\pm\sigma_2$ & $\tilde{X}$ & 
      $\sigma_{12}$ \\
    \units{MeV} & \multicolumn{3}{c}{(counts $\days^{-1}\gw^{-1}$)}  & 
              (counts $\days^{-1}\gw^{-1})^2$ \\
      \hline
      \Rule
      $1.2$ & $0.151\pm 0.031$ & $0.176\pm 0.035$ & $0.172$ &  
              $-2.2\cdot 10^{-4}$ \\
      \Rule
      $2.0$ & $0.490\pm 0.039$ & $0.510\pm 0.047$ & $0.532$ &
              $-1.5\cdot 10^{-4}$ \\
      \Rule
      $2.8$ & $0.656\pm 0.041$ & $0.610\pm 0.049$ & $0.632$ &
              $-3.5\cdot 10^{-4}$ \\
      \Rule
      $3.6$ & $0.515\pm 0.036$ & $0.528\pm 0.044$ & $0.530$ & 
              $-3.3\cdot 10^{-4}$ \\
      \Rule
      $4.4$ & $0.412\pm 0.033$ & $0.408\pm 0.040$ & $0.379$ & 
              $-2.0\cdot 10^{-4}$ \\
      \Rule
      $5.2$ & $0.248\pm 0.030$ & $0.231\pm 0.034$ & $0.208$ & 
              $-0.7\cdot 10^{-4}$ \\
      \Rule
      $6.0$ & $0.102\pm 0.023$ & $0.085\pm 0.026$ & $0.101$ & 
              $-1.3\cdot 10^{-4}$ \\
      \hline
    \end{tabular}
  \end{center}
\end{table}
%
%

\section{Neutrino oscillation tests}
Since no evidence was found for a deficit of measured vs. expected neutrino 
interactions, we can derive from the data the exclusion plots in the 
plane of the oscillation parameters $(\delta m^2,\sin^2 2 \theta)$, in 
the simple two-neutrino oscillation model.

We employed three methods, each characterised by a different dependence on 
statistical and  systematic errors and each having a different sensitivity 
to oscillations.

Analysis ``A'' \\
Experimental input: the measured positron spectra $X_1(E)$ and $X_2(E)$ 
from each reactor. Computed reference inputs: the predicted positron 
spectrum, obtained by merging the reactor information, the neutrino 
spectrum model and the detector response; the two-flavour survival 
probability . 
``A'' uses all the experimental information available; it directly depends 
on the correct determination of the integrated neutrino flux, number of 
target protons, detection efficiencies and the $\anu$ cross section.
    
Analysis ``B'' \\
Experimental input: the ratio of the measured positron spectra 
$X_1(E)$ and $X_2(E)$ from the two, different distance, reactors . 
Computed reference inputs: the two-flavour survival probability .
``B'' is almost completely independent of the correct determination of the 
integrated neutrino flux, number of target protons, detection efficiencies.
Statistical errors dominate.

Analysis ``C'' \\
Experimental input: the measured positron spectra $X_1(E)$ and $X_2(E)$ 
from each reactor. Computed reference inputs: the shape of the predicted 
positron spectrum, the absolute normalization being left free.
The only relevant systematic uncertainty comes from the precision of 
the neutrino spectrum extraction method~\cite{Sch85}.

\subsection{Results from analysis ``A''}
In the two-neutrino oscillation model, the expected positron spectrum 
$\overline{X}$ can be parametrized as follows:
\begin{equation}
  \overline{X}(E_j,L_k,\theta,\delta m^2) = \tilde{X}(E_j) 
  \overline{P}(E_j,L_k,\theta,\delta m^2),
  \quad (j = 1,\ldots,7 \quad k = 1,2)
  \label{xosc}
\end{equation}
where $\tilde{X}(E_j)$ is the previously defined positron spectrum (independent
of distance in the absence of neutrino oscillations), $L_k$ is the 
reactor-detector distance and $\overline{P}$ is the survival probability, 
averaged over the energy bin and the finite detector and reactor core sizes. 
In order to test the compatibility of a certain oscillation hypothesis 
$(\delta m^2,\sin^2 2 \theta)$ with the measurements, we must build a 
$\chi^2$ statistic containing the $7$ experimental yields for each of 
the two positions $L_k$ (listed in Table~\ref{yieboth}). We group these
values into a $14$-element array $X$, as follows:
\begin{equation}
  \vec{X} = (X_1(E_1),\ldots,X_1(E_7),X_2(E_1),\ldots,X_2(E_7)),
  \label{xarray}
\end{equation}
and similarly for the associated variances. These components are not 
independent, as yields corresponding to the same energy bin are extracted
for both reactors
simultaneously, and the off-diagonal matrix elements $\sigma_{12}$ (also listed 
in Table~\ref{yieboth}) are non-vanishing. By combining the statistical
variances with the systematic uncertainties related to the neutrino spectrum,
the $14 \times 14$ covariance matrix can be written in a compact form as 
follows:
\begin{equation}
  V_{ij} = \delta_{i,j}(\sigma_i^2 + \tilde{\sigma}_i^2) + 
           (\delta_{i,j-7} + \delta_{i,j+7})\sigma^{(i)}_{12}
              \qquad (i,j = 1,\ldots,14),
  \label{cormat}
\end{equation}
where $\sigma_i$ are the statistical errors associated with the yield array
(Eqn.\ \ref{xarray}), $\tilde{\sigma}_i$ are the
corresponding systematic uncertainties, and 
$\sigma^{(i)}_{12}$ are the statistical covariances of 
the reactor $1$ and 
$2$ yield contributions to the i-th energy bin (see Table~\ref{yieboth}). The 
systematic errors, which include the statistical error on the $\beta$-spectra 
measured at ILL~\cite{Sch85} as well as the bin-to-bin systematic error 
inherent in the conversion procedure, range from $1.4 \,\%$ at $2 \units{MeV}$ 
(positron energy) to $7.3 \,\%$ at $6 \units{MeV}$ and are assumed to be 
uncorrelated.

 We next  take into account the systematic error 
related to the absolute normalization; combining all the contributions 
listed in Table~\ref{sysnorm}, we obtain an overall normalization uncertainty 
of $\sigma_\alpha = 2.7 \,\% $.
%
\begin{table}[htb]
  \caption{\small Contributions to the overall systematic uncertainty on the 
    absolute normalization factor.}
  \label{sysnorm}
  \begin{center}
    \begin{tabular}{lc}
      \hline
      parameter                   & relative error $(\%)$ \\
      \hline
      reaction cross section      & $1.9 \,\%$ \\
      number of protons           & $0.8 \,\%$ \\
      detection efficiency        & $1.5 \,\%$ \\
      reactor power               & $0.7 \,\%$ \\
      energy absorbed per fission & $0.6 \,\%$ \\
      \hline
      combined                    & $2.7 \,\%$ \\
      \hline
    \end{tabular}
  \end{center}
\end{table}

\begin{figure}[htb]
  \begin{center}
    \mbox{\epsfig{file=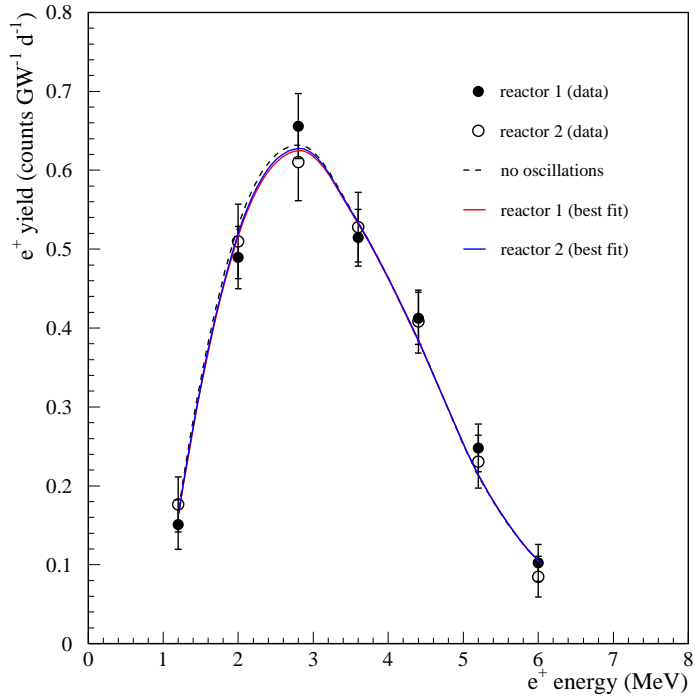,width=0.7\linewidth}}
    \caption{\small Positron spectra for reactor $1$ and $2$; the solid curves
             represent the predicted spectra corresponding to the
             analysis A 
       best-fit parameters, the dashed one to that predicted  for 
              no oscillations.}
    \label{yiemin}
  \end{center}

\end{figure}

We now define
\begin{eqnarray}
  \chi^2(\theta,\delta m^2,\alpha,g) & = &\sum_{i=1}^{14} \sum_{j=1}^{14}
  (X_i - \alpha \overline{X}(g E_i,L_i,\theta,\delta m^2)) V_{ij}^{-1} 
  (X_j - \alpha \overline{X}(g E_j,L_j,\theta,\delta m^2)) + \nonumber\\
  & + &\big( \frac{\alpha-1}{\sigma_\alpha} \big)^2 + 
  \big( \frac{g-1}{\sigma_g}
  \big)^2,
  \label{chiA}
\end{eqnarray}
%
%
where $\alpha$ is the absolute normalization constant and $g$ is the 
energy-scale calibration factor. The  uncertainty  in $g$ is 
$1.1 \,\%$, resulting from the accuracy on the energy scale 
calibration ($16 \units{KeV}$ at the $2.11 \units{MeV}$ visible energy line 
associated with the n-capture on Hydrogen) and the $0.8 \,\%$ drift in the 
Gd-capture line, as measured throughout the acquisition period with 
high-energy spallation neutrons (see Fig.~\ref{stab}). 
The function (Eqn.~\ref{chiA}) is a $\chi^2$ with 12 degrees of freedom.
The minimum value $\chi^2_{min} = 5.0$ (corresponding to a 
$\chi^2$ probability $P_{\chi^{2}} = 96 \,\%$) is found for the parameters 
$\sin^2 2 \theta = 0.23$, $\delta m^2 = 8.1\cdot 10^{-4}\units{eV^2}$, 
$\alpha = 1.012$, $g = 1.006$. The resulting positron spectra are shown 
by solid lines in Fig.~\ref{yiemin} superimposed on the data.
Also the no-oscillation hypothesis, with $\chi^2(0,0) = 5.5$, 
$\alpha = 1.001$ and $g = 1.006$, is found to be in excellent 
agreement with the data ($P_{\chi^{2}} = 93 \,\%$).

\begin{figure}[hp]
  \begin{center}
    \mbox{\epsfig{file=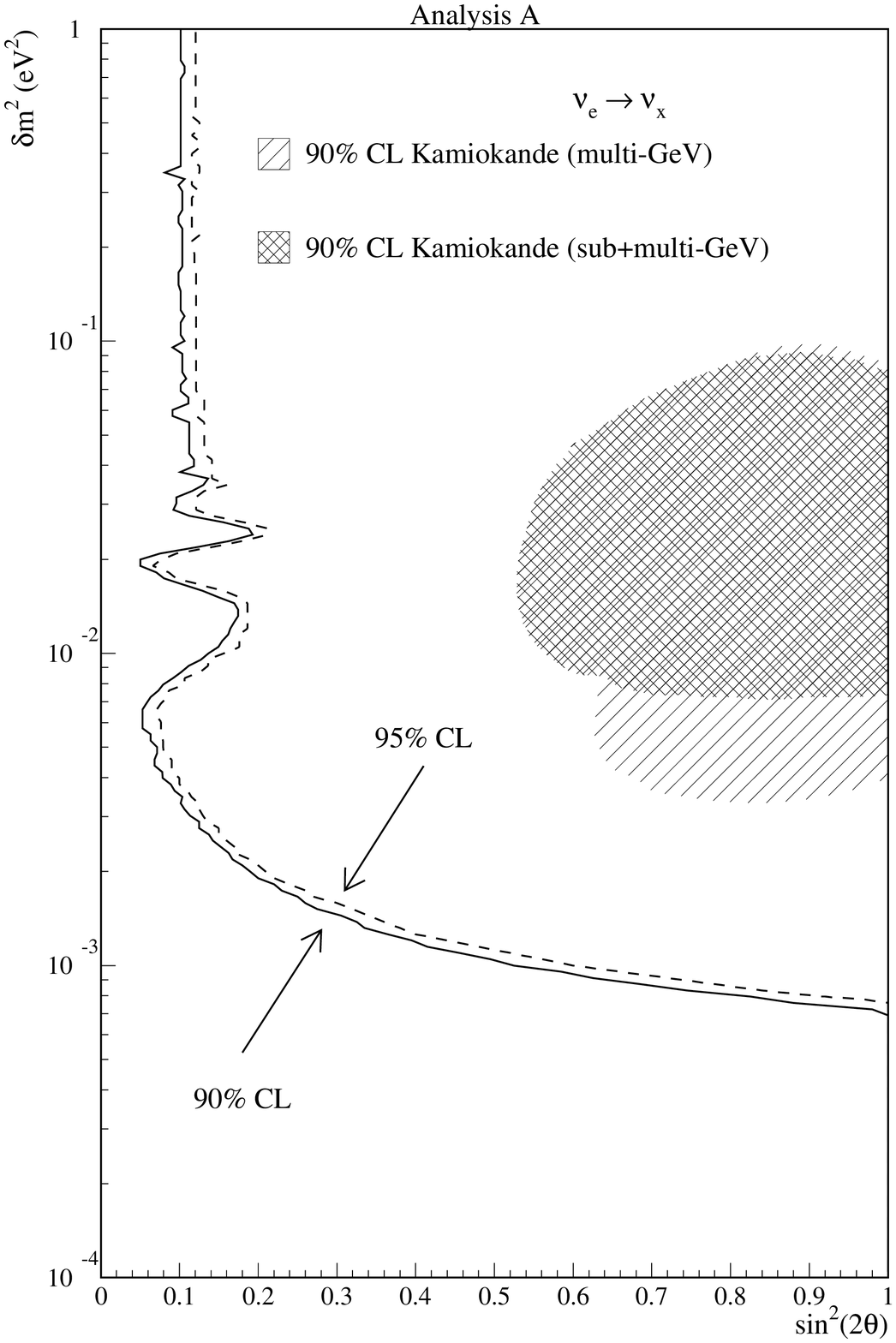,width=0.9\linewidth}}
    \caption{\small Exclusion plot for the oscillation parameters based on the 
      absolute comparison of measured vs. expected positron yields.}
    \label{exclA}
  \end{center}
\end{figure}

To test a particular oscillation hypothesis $(\delta m^2,\sin^2 2 \theta)$ 
against the parameters of the best fit, we adopted the 
Feldman \& Cousins prescription~\cite{Fel98}. 
The exclusion plots at the $90 \,\%$~C.L. (solid line) and $95 \,\%$~C.L. 
are shown in Fig.~\ref{exclA}.  
The region allowed by Kamiokande\cite{Kam} for the $\numu \rightarrow \nue$ 
oscillations is also shown for comparison. The $\delta m^2$ limit at 
full mixing is $7 \cdot 10^{-4} \units{eV^2}$, to be compared with 
$9.5 \cdot 10^{-4} \units{eV^2}$ previously published\cite{Apo98}. 
The limit for the mixing angle in the asymptotic range of large mass 
differences is $\sin^2 2 \theta = 0.10$, which is better by   a factor of
two than
the previously published value (as recalculated according to~\cite{Fel98}) .
%

\subsection{Results from analysis ``B''}
\begin{figure}[hp]
  \begin{center}
    \mbox{\epsfig{file=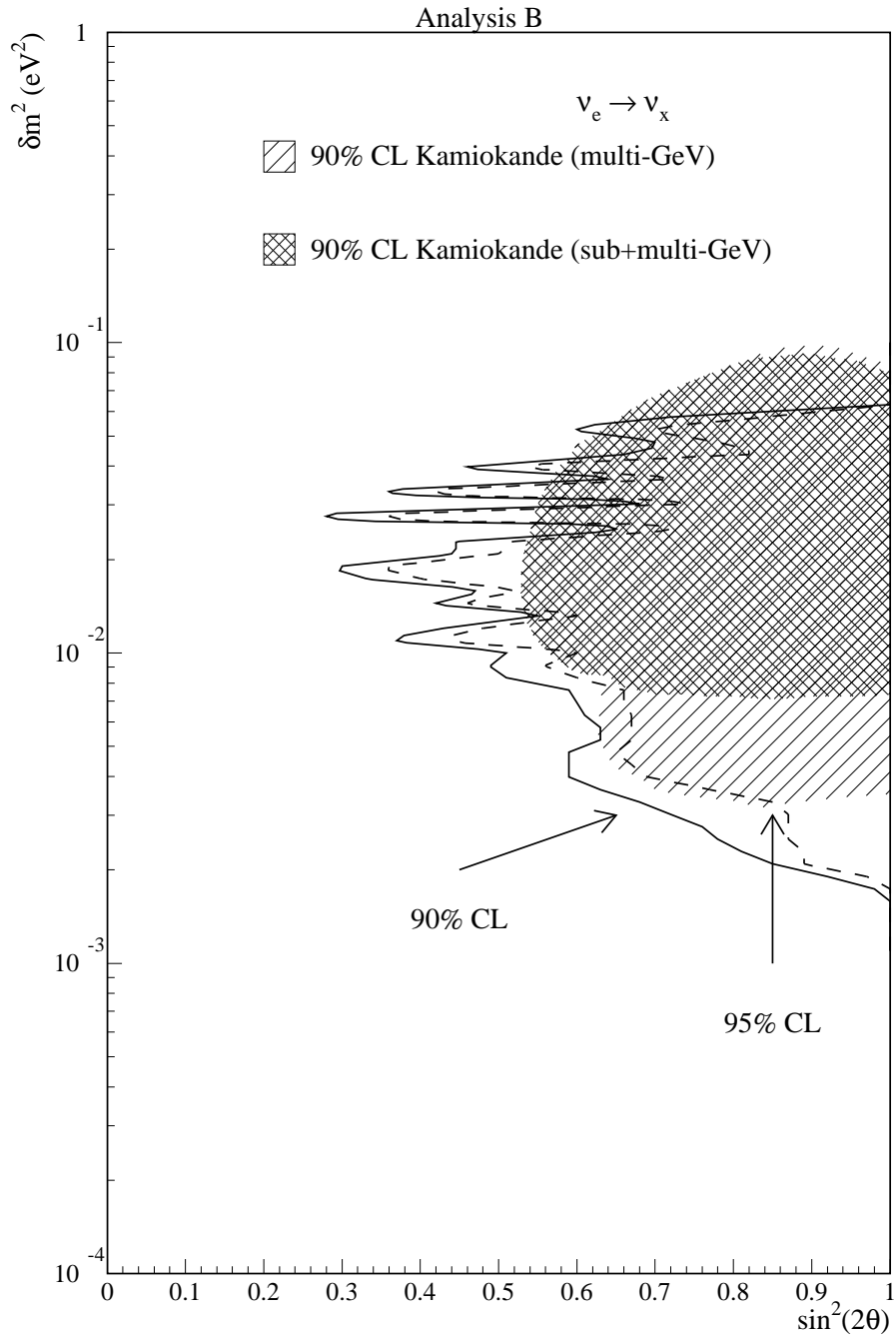,width=0.9\linewidth}}
    \caption{\small Exclusion plot contours at $90 \,\%$~C.L. and 
      $95 \,\%$~C.L. obtained from the ratios of the positron spectra 
      from the two reactors.}
    \label{exclplotB1}
  \end{center}
\end{figure}

The ratio $R(E_i)\equiv X_1(E_i)/X_2(E_i)$ of the measured positron spectra 
is compared with its expected values. Since the expected spectra are 
the same for both reactors in the case of no-oscillations, the expected ratio 
reduces to the ratio of the average survival probabilities in each 
energy bin. We can then form the following $\chi^2$ function:
\begin{equation}
 \chi^2 = \sum_{i=1}^7 \big( \frac{R(E_i)-\overline{R}
    (E_i,\theta,\delta m^2)}{\delta R(E_i)} \big)^2
 \label{chiB1}
\end{equation}
where $\delta R(E_i)$ is the statistical uncertainty on the measured ratio. 
We adopted the same procedure described in the previous section to determine 
the confidence domain in the $(\delta m^2,\sin^2 2 \theta)$ plane. 
The resulting exclusion plot is shown in Fig.~\ref{exclplotB1}; the 
contour lines of the $90 \,\%$ and $95 \,\%$~C.L. are drawn. 
Although less powerful than analysis ``A'', the region excluded by this 
oscillation test nevertheless almost completely covers the one allowed by Kamiokande.
%
\subsection{Results from analysis ``C''}
\begin{figure}[htb]
  \begin{center}
    \mbox{\epsfig{file=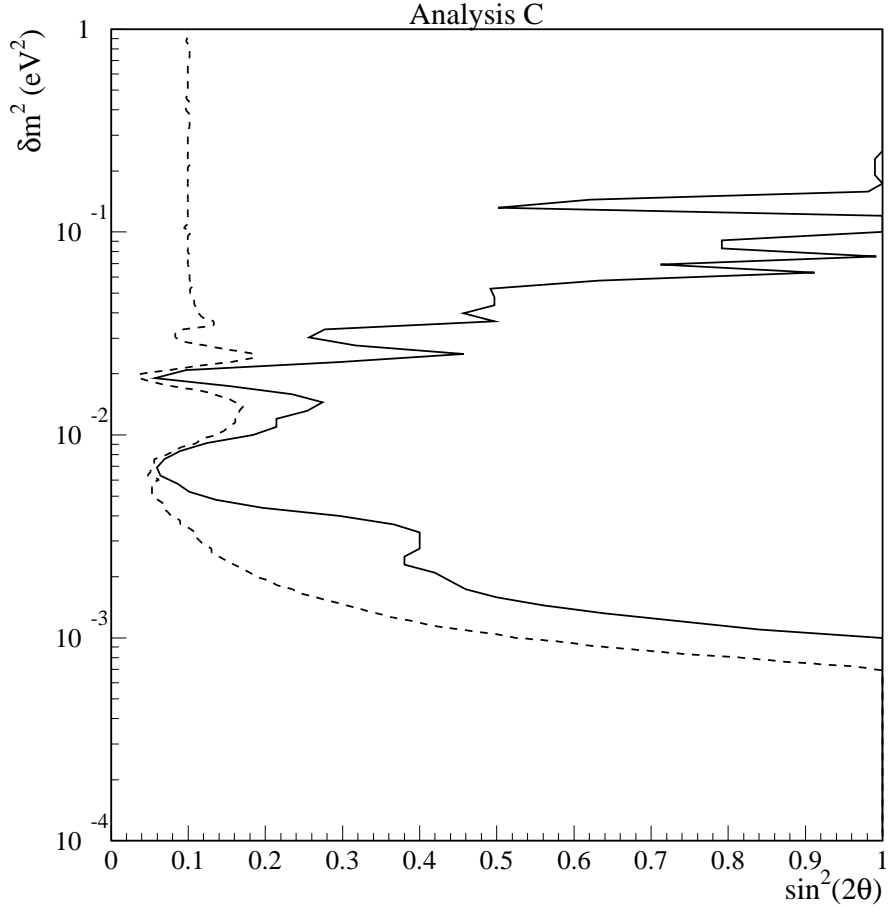,width=0.9\linewidth}}
    \caption{\small Exclusion plot at $90 \,\%$~C.L. obtained by the 
      shape test (analysis C, solid line)  compared to the integral 
      test (analysis A, dashed line).}
    \label{exclplotC}
  \end{center}
\end{figure}

Analysis ``C'' is mathematically similar to analysis ``A'', the only 
difference being the omission of  the absolute normalization; in ``A'' we forced  
the integral counting rate to be distributed around the predicted value 
($\alpha = 1$), with a $\sigma_\alpha = 2.7 \,\%$ systematic uncertainty; 
in ``C'', $\alpha$ is left free (which is equivalent to 
$\sigma_\alpha = \infty$).
%
%
\begin{equation}
 \begin{array}{ll}
  \chi^2(\theta,\delta m^2,\alpha,g) ~ = & \nonumber \\
                                         & \nonumber \\ 
  \sum_{i=1}^{14} \sum_{j=1}^{14}
  (X_i - \alpha \overline{X}(g E_i,L_i,\theta,\delta m^2)) V_{ij}^{-1} 
  (X_j - \alpha \overline{X}(g E_j,L_j,\theta,\delta m^2)) ~ + 
                                         & \nonumber \\
                                         & \nonumber \\
  + ~ \big( \frac{g-1}{\sigma_g} \big)^2 &
  \label{chiC}
 \end{array}
\end{equation}
The exclusion plot, obtained according to the Feldman-Cousins
prescriptions, is shown in Fig.~\ref{exclplotC} and compared to the results 
of analysis ``A''.
%
\section{Conclusions}
Since publishing its initial findings, 
the CHOOZ experiment has considerably improved both its statistics and the 
understanding of systematic effects. As a result it finds, at 90 \,\% C.L., 
no evidence for neutrino oscillations in the disappearance mode
$\overline{\nu}_{e} \rightarrow \overline{\nu}_{x}$ for the parameter 
region given by approximately  $ \dmsq > 7 \cdot 10^{-4}\units{eV^2}$ for 
maximum mixing, and $\sinsq = 0.10$ for large $\dmsq$, as shown in 
Fig.~\ref{exclA}. A lower sensitivity result, but independent of most 
of the systematic effects, is able, alone, to almost completely exclude the 
Kamiokande allowed oscillation region. 
\section{Acknowledgements}
We wish to thank Prof.~Gianni Fiorentini, for  
initial and fruitful discussions on the two-reactor comparison. \\
We thank Prof.~Erno Pretsch and his group at ETH Zurich, for some 
precise measurements of the target scintillator hydrogen content. \\
Construction of the laboratory was funded by \'Electricit\'e de France
(EdF). Other work was supported in part by IN2P3--CNRS (France), INFN
(Italy), the United States Department of Energy, and by RFBR (Russia).
We are very grateful to the Conseil G\'en\'eral des Ardennes for having
provided us with the headquarters building for the experiment. At
various stages during construction and running of the experiment, we
benefited from the efficient work of personnel from SENA (Soci\'et\'e
Electronucl\'eaire des Ardennes) and from the EdF Chooz B nuclear plant.
Special thanks to the technical staff of our laboratories for their
excellent work in designing and building the detector.

%
\end{document}